\documentclass[aps,prb,amsmath,amssymb,twocolumn,superscriptaddress,showpacs,floatfix]{revtex4-1}
\usepackage{ifpdf}
 \newif\ifpdf
\ifx\pdfoutput\undefined
   \pdffalse
\else
   \pdfoutput=1
   \pdftrue
\fi
\ifpdf
   \usepackage{graphicx}
   \usepackage{epstopdf}
\else
   \usepackage{graphicx}
\fi

\usepackage{bm}
\usepackage{epsfig}
\usepackage[usenames]{color}
\usepackage{soul}
\usepackage{graphicx}
\newcommand{\Cs}{\mathit C_{\scriptscriptstyle{\Sigma}}}
\newcommand{\Rs}{\mathit R_{\scriptscriptstyle{\Sigma}}}

\begin{document}

\title[Interplay of ferroelectricity and single electron tunneling]{Interplay of ferroelectricity and single electron tunneling}

\author{S.~A.~Fedorov}
 \affiliation{Department of Theoretical Physics, Moscow Institute of Physics and Technology, Moscow 141700, Russia}
 \affiliation{P.N. Lebedev Physical Institute of the Russian Academy of Sciences, Moscow 119991, Russia}
\author{A.~E.~Korolkov}
\affiliation{Department of Theoretical Physics, Moscow Institute of Physics and Technology, Moscow 141700, Russia}
\affiliation{P.N. Lebedev Physical Institute of the Russian Academy of Sciences, Moscow 119991, Russia}
\author{N.~M.~Chtchelkatchev}
\affiliation{Department of Theoretical Physics, Moscow Institute of Physics and Technology, Moscow 141700, Russia}
\affiliation{Department of Physics and Astronomy, California State University Northridge, Northridge, CA 91330, USA}
\affiliation{L.D. Landau Institute for Theoretical Physics, Russian Academy of Sciences,117940 Moscow, Russia}
\affiliation{Institute for High Pressure Physics, Russian Academy of Sciences, 142190, Moscow, Russia}
\author{O.~G.~Udalov}
\affiliation{Department of Physics and Astronomy, California State University Northridge, Northridge, CA 91330, USA}
\affiliation{Institute for Physics of Microstructures, Russian Academy of Science, Nizhny Novgorod, 603950, Russia}
\author{I.~S.~Beloborodov}
\affiliation{Department of Physics and Astronomy, California State University Northridge, Northridge, CA 91330, USA}

\date{\today}

\begin{abstract}
We investigate the interplay of ferroelectricity and quantum electron transport at the nanoscale in the regime of Coulomb blockade. Ferroelectric polarization in this case is no longer the external parameter but should be self-consistently calculated along with electron hopping probabilities leading to new physical transport phenomena studying in this paper. These phenomena appear mostly due to effective screening of a grain electric field by ferroelectric environment rather than due to polarization dependent tunneling probabilities. At small bias voltages polarization can be switched by a single excess electron in the grain. In this case transport properties of SET exhibit the instability (memory effect).
\end{abstract}

\maketitle

Systems with ferroelectric (FE) elements attract much of attention due to their interesting fundamental properties at the nanoscale as well as due to their possible applications
in microelectronics, especially in nonvolatile memory devices, in emerging technologies of Terahertz-detecting and in building of advanced (nano)capacitors.~\cite{dawber2003self,ahn2004ferroelectricity,Dawber2005RevModPhys,wang2007modeling,zhang2007improved,Scott2007,Maksymovych2009Science,chu2008Nature,lee2008nanocapacitor,kalinin2010RPP,Dawber2012,Dawber2012_1,ortega2012relaxor,Chanthbouala2012}
In quantum junctions the ferroelectricity influences electron transport: Tunneling through the FE barriers shows giant electro-resistance effect caused by the strong dependence of electron tunneling probability on the FE polarization and external bias orientations.~\cite{Maksymovych2009Science,Tsy2010} Here we focus on the inverse process --- the influence of electron transport on ferroelectricity.~\cite{ahn2004ferroelectricity,kalinin2010RPP} The naive guess would be that a single electron, small quantum object, can slightly influence the macroscopic effect --- ferroelectricity. However, we show that this is not quite true and discuss the interplay of ferroelectricity and quantum electron transport at the nanoscale in the regime of Coulomb blockade. Polarization in this case is no longer the external parameter but should be self-consistently calculated along with electron hopping probabilities leading to new physical transport phenomena studying in this paper. These phenomena appear mostly due to effective screening of a grain electric field by ferroelectric environment rather than due to polarization dependent tunneling probabilities.

Ferroelectrics (FE) are characterized by the polarization $\mathbf P$ whose direction and magnitude can be changed by applying an external electric field $\bf\mathcal{E}$ larger than the ferroelectric switching field, $\mathcal E_s$. The ground ferroelectric state of a bulk sample is usually not uniformly polarized but divided into domains to lower the electrostatic energy, like in ferromagnets.~\cite{landau2004electrodynamics}

\begin{figure}[t]
  \centering
  \includegraphics[width=0.9\columnwidth]{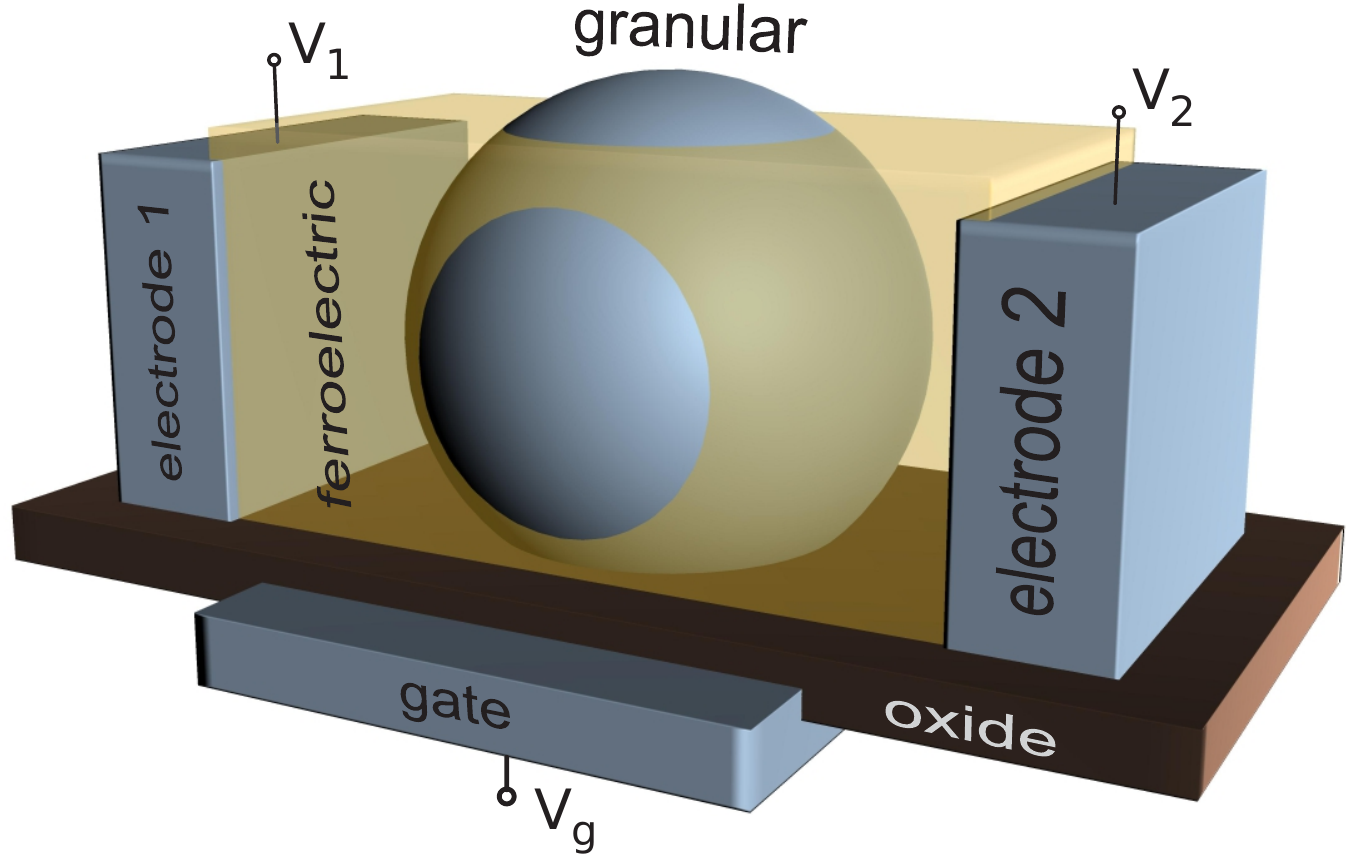}\\
  \caption{(Color online) Sketch of a single electron device with ferroelectric tunnel junctions.}\label{fig1}
\end{figure}
At the nanoscale to influence the polarization of (nano)ferroelectric one can apply strong enough bias to nanotips:~\cite{ahn2004ferroelectricity} There is a well developed technique of imaging and control of domain structures in ferroelectric thin films by a tip of a scanning probe microscope, see, e.g., Refs.~\onlinecite{gruverman1998rev,ahn2004ferroelectricity,Kalinin2008APL,rodriguez2009NanoLett,Maksymovych2009Science,kalinin2010RPP}.

Here we show how ferroelectric polarization switching can be produced by placing a single excess electron at the nanograin. Charged metal particle creates a strong enough electric field, $\mathcal E\approx 1$~MV/cm around it.
Numerous ferroelectric (nano)materials have the same order of magnitude switching
field.~\cite{ahn2004ferroelectricity,kalinin2010RPP}

We study a single electron device with electric current flowing from the source to the drain electrodes with  voltages $V_1$ and $V_2$, respectively, Fig.~\ref{fig1}. A metallic nanoparticle is placed in between these electrodes. The third gate-electrode controls the effective number of electrons on the grain through the capacitive coupling. We assume that the charging energy $E_c$ of a single grain is the leading energy scale in the problem, $E_c \gg T$ with $T$ being the temperature. The device shown in Fig.~\ref{fig1} is a standard Single Electron Transistor (SET)~\cite{averin1991single,averin1991theory,devoret1992single,wasshuber2001computational,Glatz2012PRB86Nonlinear,Chtchelkatchev2013JPhys,Chtchelkatchev2013PRB88Universality,Kafanov2013JAP} with one important exception: electrons tunnel through ferroelectric insulating layers.

\begin{figure}[t]
  \centering
  \includegraphics[width=0.8\columnwidth]{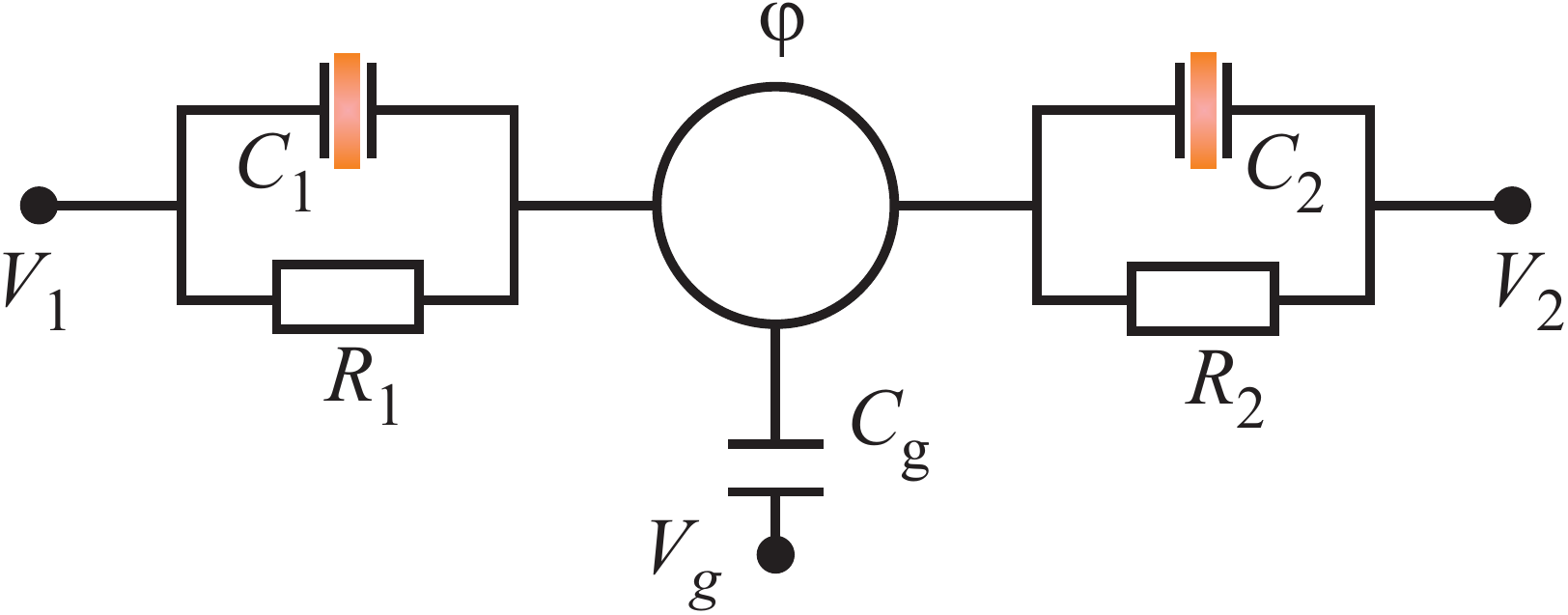}  \\
  \caption{(Color online) Effective circuit equivalent to the setup shown in Fig.~\ref{fig1}. Ferroelectric insulators are highlighted by the orange-color.  }\label{fig2}
\end{figure}

The tunnel junctions between the nanograin and the electrodes form the capacitors with ferroelectric filling (see equivalent electric circuit Fig.~\ref{fig2}). Typically, ferroelectric placed into the capacitor chooses polarization direction perpendicular to the electrodes. This configuration reduces electrostatic energy due to FE polarization screening by the electrodes. The direction of polarization can be switched applying the bias voltage to the capacitor. In SET the potentials of the electrodes and the gate potential are usually fixed. The grain potential $\phi$ can fluctuate and can be found by solving simultaneously the electrostatic and the electron transport problems. The potential $\phi$ depends not only on the bias voltage and capacitances, but also on the probability
distribution $p(n)$ to find $n$ electrons on the grain and on the polarization of ferroelectrics. Polarizations of ferroelectric layers in turn depend on the grain potential and $p(n)$. Thus we need to consider the self-consistent problem.

The solution of self-consistent problem strongly depends on the relaxation parameters of ferroelectric material: How quickly the polarizations can change (flip) during the characteristic time of charging(discharging) of SET by a single electron. Below we focus on two limiting cases when both ferroelectric layers have relaxation times  much longer than one-electron charging-discharging time and vice-versa. These two cases correspond to qualitatively different behavior of FE SET.

\begin{figure}[t]
\begin{center}
\includegraphics[width=1\columnwidth, keepaspectratio]{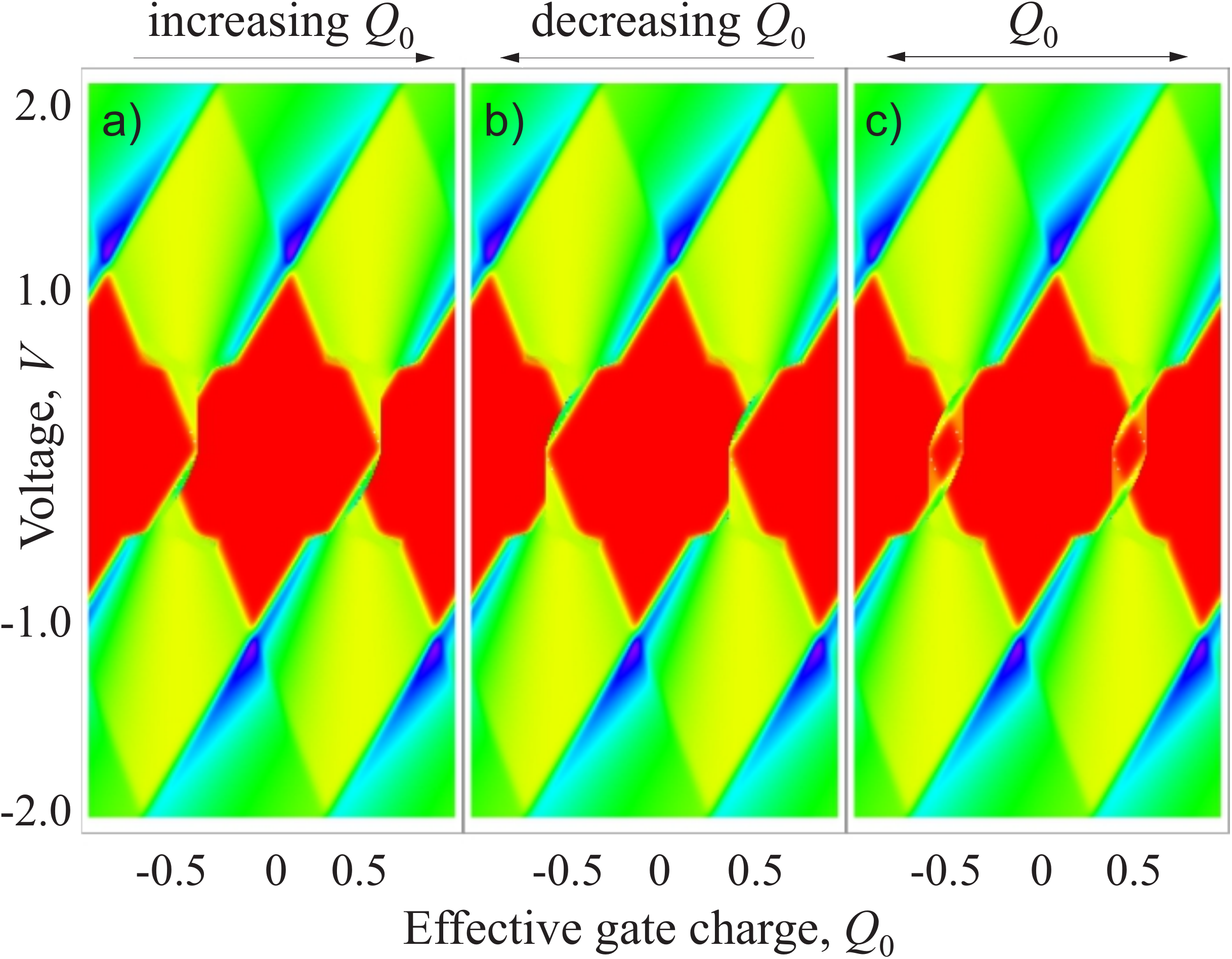}
\caption{(Color online) Coulomb Diamonds --- the conductance density plot. Here $V_{1,2}=\mp V/2$.
Graphs (a) and (b) differ by the change of the evolution direction of parameter
$Q_0=-C_gV_g$.
Graph (c) is shown for forward-backward evolution of parameter $Q_0$. The dimensionless temperature is
$T=0.01$ and all other parameters are similar to Fig.~\ref{fig4}.}
\label{fig:Diamonds}
\end{center}
\end{figure}

\begin{figure*}[t]
\begin{center}
\includegraphics[width=1\textwidth, keepaspectratio]{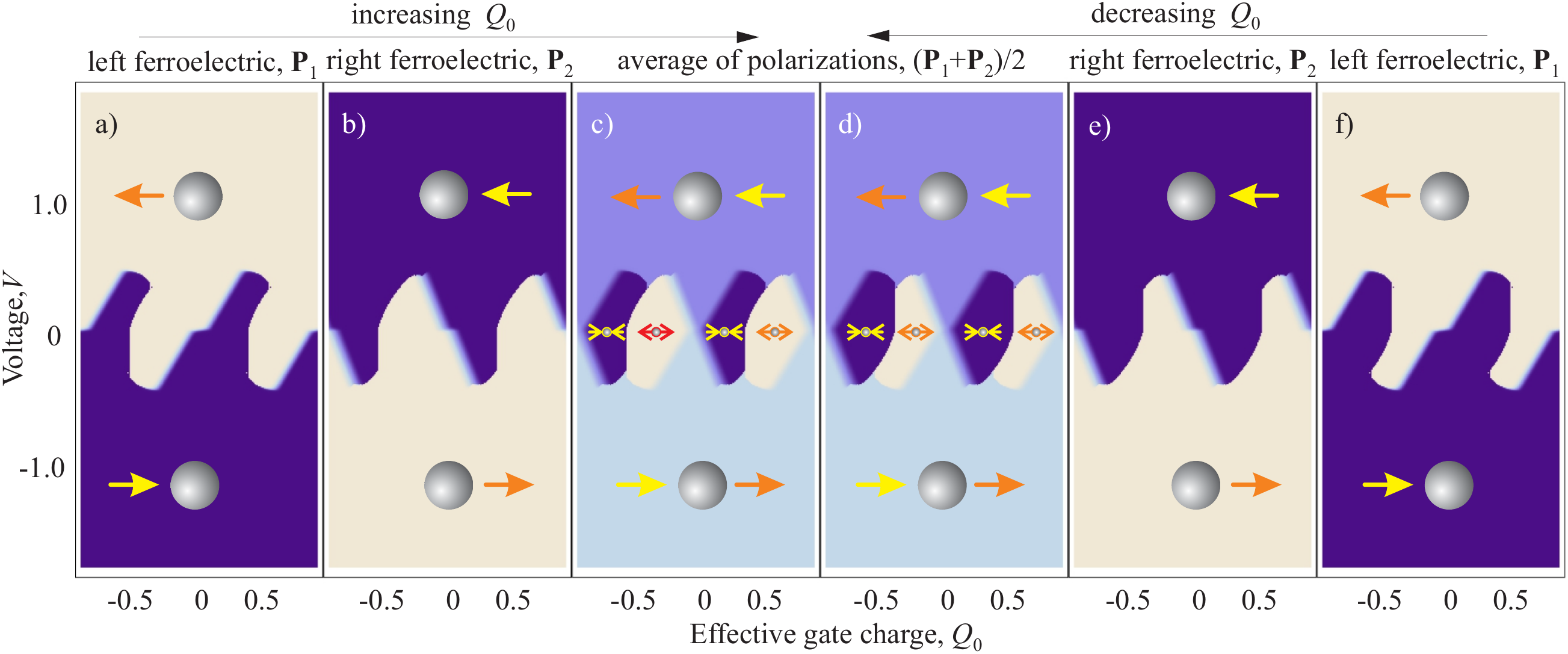}
\caption{(Color online) The density plot shows the ``phase diagram'' of the ferroelectric SET in $(Q_0,V)$-space, where
color gradients stand for polarization. The arrows show the polarizations of the left and the
right ferroelectrics. Graphs (a) and (b) show ferroelectric polarizations for increasing parameter
$Q_0$ while graphs (e) and (f) correspond to the decreasing $Q_0$. Graphs (c) and (d) show the arithmetical mean of the polarizations corresponding to the left and the right ferroelectrics. All parameters are similar to Fig.\ref{fig:Diamonds}
except the parameter $q_i^0$: $q_1^0=0.03$, and $q_2^0=0.06$ in Eq.~\eqref{qfe}.}
\label{figPhaseDiag}
\end{center}
\end{figure*}

The case of slow FE is considered in Sec.~\ref{Sec1}. We study
the dependence of FE state on bias and gate voltages and show
that the Coulomb diamonds have the ``fine-structure'' mediated by ferroelectricity
that depends on the gate-voltage, Fig.~\ref{fig:Diamonds} [at large enough ferroelectric polarizations this fine-structure can become comparable with the size of the diamonds]. We present the plot of FE ``phase diagram'', Fig.~\ref{figPhaseDiag}. For large bias voltages polarization in both capacitors are co-directed and does not affect the electron transport. At small bias voltages the polarization can be switched by a single excess electron in the grain. In this case transport properties of SET exhibit the instability (hysteresis), Fig.~\ref{fig:Diamonds}.
We emphasize that this instability appears even without the hysteresis of polarization $P(\mathcal E)$.

In Sec.~\ref{Sec2} we discuss the case of fast FE. Then the instability is absent.
However, we show that the Coulomb-Blockade peaks of zero-bias conductance as the function of the gate-voltage~\cite{averin1991single,averin1991theory,devoret1992single,wasshuber2001computational} become wider and finally disappear with increasing of the FE polarizations. Such an effect appears due to strong non-linear screening of electron charge in the grain by ferroelectrics leading to the suppression of the Coulomb Blockade.
In Sec.~\ref{sec:disc} we discuss relation of our theory to real experimental situation.

\section{Single electron device with ferroelectric tunnel junctions}

Below we discuss the basic properties of SET sketched in Fig.~\ref{fig1}. The equivalent
electric circuit is shown in Fig.~\ref{fig2}. The ferroelectricity influences the properties of
SET through two capacitors with FE insulating layers and results in the redistribution
of charge over the surface of the nanoparticle. In particular, ferroelectric with
polarization $\mathbf P$ induces the local charge on the nanoparticle surface with
the surface density $\mathbf P\cdot \mathbf n$,~\cite{landau2004electrodynamics} where
$\mathbf n$ is the normal to the surface. The excess charge on the nano-grain is given
by the following expression
\begin{gather}\label{eq:ne}
ne = \sum_i\left \{C_i\left[\phi(n)-V_i\right]+\int_i d\mathbf n_i\cdot \mathbf P_i\right \},
\end{gather}
where $n$ is the number of excess charges, $e$ is the electron charge, $\phi(n)$ is the potential
of the nano-grain, $C_i$ with $i=1,2,g$ is the capacitance.
The surface integration is performed over the nanoparticle sides playing the
role of the capacitor plates in Fig.~\ref{fig2}.

We study the SET with fixed electrodes and gate potentials and find the grain potential $\phi(n)$
using Eq.~\eqref{eq:ne}. Following the ``orthodox model''~\cite{averin1991single,averin1991theory,devoret1992single,wasshuber2001computational} we obtain the probabilities $p(n)$ to find $n$ electrons on the grain.
In the stationary case they satisfy the detailed balance equation
\begin{equation} \label{eq:ballance}
    p(n)\Gamma_{n\to n+1}=p(n+1)\Gamma_{n+1\to n},
\end{equation}
where the transition rate $\Gamma_{n\to n+1}(\langle\phi\rangle)$ describes the change of grain charge from $n$ to $n+1$ electrons, see Appendix~\ref{Ap1}.  Calculating transition rates $\Gamma$ we neglect
the dependence of electron tunneling amplitudes on the FE orientation, however this effect can be easy included
in our consideration.
Our estimates show that consideration of polarization dependent tunneling
probabilities does not destroy the effect but it rather enhances it.

The electric current can be written in terms of the transition rates as follows
\begin{align}
    I &= e\sum_{n=-\infty}^\infty p(n) \left[ \Gamma^{(1)}_{n\to n-1}-\Gamma^{(1)}_{n\to n+1} \right] = \notag \\
    &= e\sum_{n=-\infty}^\infty p(n) \left[ \Gamma^{(2)}_{n\to n+1}-\Gamma^{(2)}_{n\to n-1} \right]. \label{eq:I}
\end{align}
Here the upper index of $\Gamma$ refers to the particular tunnel junction, see Appendix~\ref{Ap1}. Solving Eqs.~\eqref{eq:ne}-\eqref{eq:I} self-consistently we find the current $I$.

The polarization $\mathbf P$ of the FE is sensitive to the electric field
and can be flipped by strong enough field. The characteristic time scale for electron
tunneling is $\tau_e=\Rs\Cs$,
with $\Cs=\sum_i C_i$ and $\Rs=R_1+R_2$ being the total capacitance and the total resistance, respectively.
The characteristic time scale for polarization
change, $\tau_{\rm\scriptscriptstyle P}$, can be either larger or smaller than $\tau_e$.
Both cases are relevant for experiment and will be discussed below.

Here we consider the following model describing the electric
field dependence of polarization~\cite{zhang2007improved,yoon2003reduced}
\begin{gather}\label{P-approx}
P(\mathcal E)= P^{0}\tanh\left(\frac{\mathcal E}{\mathcal E_{s}}\right),
\end{gather}
where  $\mathcal E_{s}$ being a material dependent parameter. Similar dependence of polarization $P$
on the capacitor voltage has the form
$P(V)= P^{0}\tanh(V/V_{s})$, where $V_s=\mathcal E_s d$ and $d$ is the distance between the electrodes of the capacitor.
Equation~(\ref{P-approx}) describes the saturation of $P$ for large electric fields
and it results in constant electric susceptibility $\chi_e=P^{0}/\mathcal E_{s}$ for
small electric fields, $\mathcal E \ll \mathcal E_s$.
Equation~(\ref{P-approx}) neglects the spontaneous polarization and the hysteresis behavior of $P(\mathcal E)$.
This simplification is valid for FE with small switching field in comparison with
the field created by the charged grain, Sec.~\ref{sec:disc}.
Below we show that even in the absence of FE hysteresis the
SET conductance has history dependence.
To highlight this result we neglect the FE hysteresis in our consideration.
The presence of memory effect in the behavior of polarization $P(\mathcal E)$ would
add an additional hysteresis in the transport properties of SET.

\subsection{Units for numerical calculations. \label{secUnits}}

We use dimensionless units in our numerical calculations: $2E_c=e^2/\Cs$ is the unit of
energy and temperature ($k_B=1$). All charges are measured in units of elementary
charge $e$, in this units the electron has charge $-1$.
The capacitance unit is $e^2/2E_c$, thus $\Cs=1$. We choose the bare tunnel resistance of the first tunnel
junction, $R_1$, between the left electrode and the nanograin for units of tunnel
resistance, Figs.~\ref{fig1}-\ref{fig2}. Thus the unit of conductance $G$ is $1/R_1$.

\subsection{Mean-field approximation: Fast charging (discharging) and slow relaxation of polarization. \label{Sec1}}

Here we consider the limit of fast grain charging and slow relaxation of polarization, $\tau_{\rm\scriptscriptstyle P}\gg \tau_e$.
In this case the polarizations of the FE layers
are defined by the \textit{average} biases across the capacitors.
The average grain potential is given by the following expression:
\begin{gather}\label{eq_phi_av}
  \langle\phi\rangle=\sum_{n=-\infty}^\infty p(n) \phi(n).
\end{gather}
Below we show that $\langle \phi\rangle$ and $p(n)$ depend on the polarization of the FE
layers that in turn depends on the average potential
$\langle \phi\rangle$ leading to the self-consistent problem.

We choose $V_1 = -V/2$ and $V_2 = V/2$ for the biases applied to electrodes, solve Eq.~\eqref{eq:ne} for the grain potential along with Eq.~\eqref{P-approx} and find
\begin{gather}\label{eqphia}
 \phi(n)=\frac{e}{C_{\Sigma}}\left\{ne-\left[Q_0+q_{\rm fe}+(C_1-C_2)\frac{V}{2}\right]\right\},
 \\\label{qfe}
 q_{\rm fe} =q^0_1\tanh\left(\frac{\langle\phi\rangle+\frac{V}{2}}{V_s}\right)+ q^0_2\tanh\left(\frac{\langle\phi\rangle-\frac{V}{2}}{V_s}\right),
\end{gather}
where $Q_0 = -C_g V_g$, $q^0_i=P^0_i S_i$ with $i=1,2$ and $S_i$ being the effective capacitance area.
We notice that parameter $q^0_i$ is positive. Comparing Eq.~(\ref{eqphia}) with the orthodox theory of SET~\cite{devoret1992single,wasshuber2001computational} we find that
the presence of ferroelectricity shifts the ``gate charge'' $Q_0$ by
the polarization-dependent constant, $q_{\rm fe}$, see Appendix~\ref{Ap1}.

We start our consideration with approximate solution of Eq.~\eqref{eq_phi_av}.
When the current flows through the ferroelectric SET the induced FE
charge stays the same. Therefore if we assume that the sum of the effective charges
induced by the FE on the grain, $q_{\rm fe}$ is known we can calculate
the probability distribution of $ n $ electrons using the orthodox theory
of SET. The only difference between the orthodox theory and our case is
the presence of an additional shift in the parameter $Q_0$.

We assume the following: a) The induced FE charges are much smaller than the
electron charge, $|q_{\rm fe} | \ll | e |$ and b) The bias voltage $V$ between the
first and the second electrodes of the transistor is much smaller
than the charging energy, $eV \ll E_c$.
For $ (Q_0 / e - 1/2) \ll 1$ and $ (Q_0 + q_{\rm fe}) / e - 1/2 \ll 1$ only zero or one
excess electron can be found on the grain with appreciable probability which
can be obtained using the orthodox theory, Appendix~\ref{Ap2}
\begin{multline}\label{eqphiap}
\frac{e\langle\phi\rangle}{E_c}=\tanh\left(\frac{E_c}{T}\frac{ \left[\delta Q_0+q_{\rm fe}\left(\langle\phi\rangle\right)\right]}{e}\right)-
\\
2\frac{\delta Q_0+q_{\rm fe}\left(\langle\phi\rangle\right)}{e}.
\end{multline}
\begin{figure}[t]
\begin{center}
\includegraphics[width=1\columnwidth, keepaspectratio]{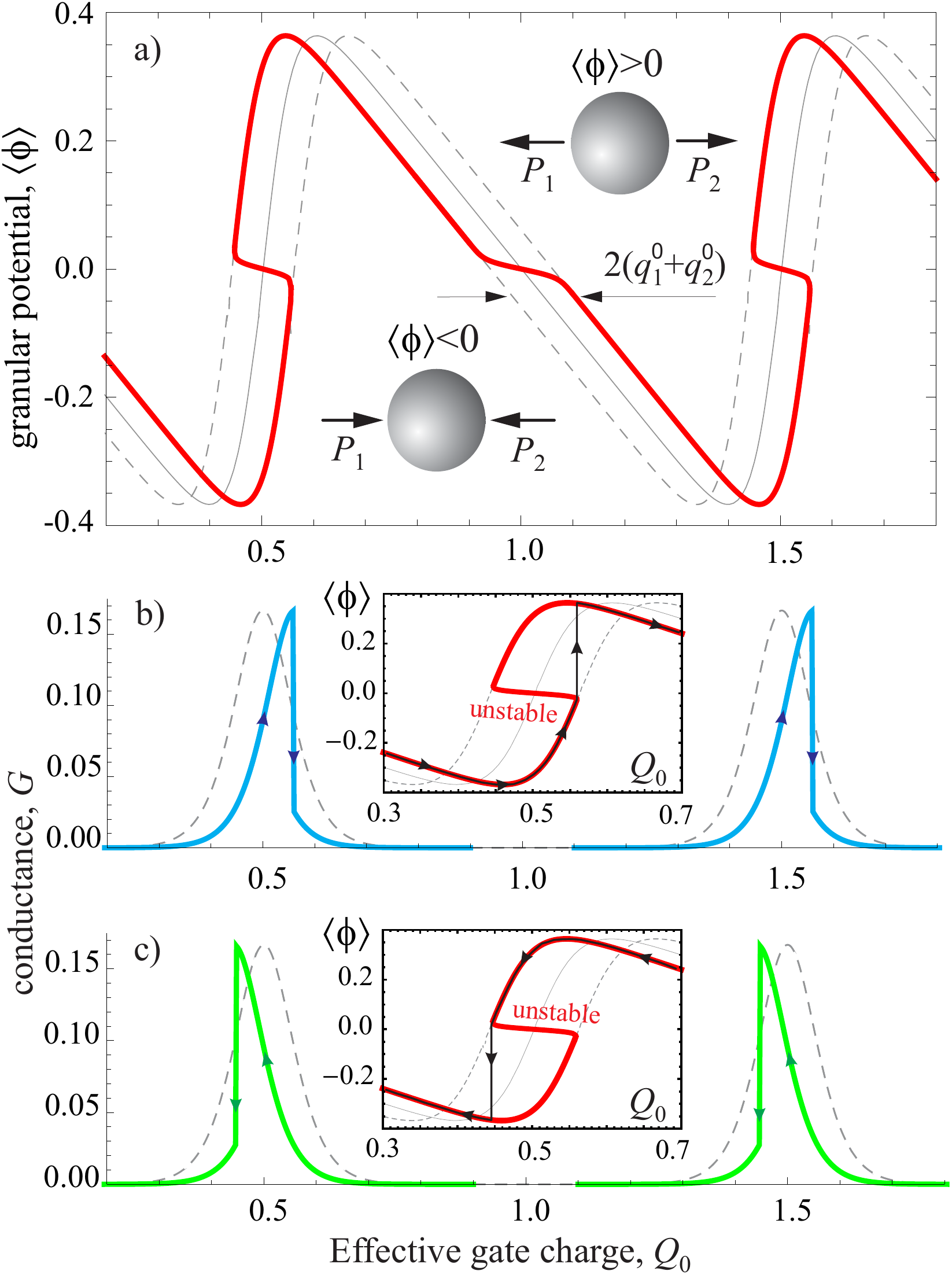}
\caption{(Color online) a) Average grain potential $\langle\phi\rangle$ for voltage $V\to 0$ vs. parameter
$Q_0$. For negative potential, $\langle\phi\rangle < 0$, polarizations of both ferroelectrics are
directed towards the nanoparticle, while for positive potential, $\langle\phi\rangle>0$, they have
the opposite direction. Plots are shown for the following set of parameters:
$q^0_1=q^0_2=0.03$, $T = 0.03$, $C_1 = 0.3$, $C_2 = 0.5$, $C_g = 0.2$, and $R_2=2R_1$.
Dimensionless units are defined in Sec.~\ref{secUnits}.
Almost linear branches of potential $\langle\phi\rangle$ with
width $2(q_1^0+q_2^0)$ correspond to the electric fields of both capacitors
smaller than the field $\mathcal E_s^{(1,2)}$ in Eq.~\eqref{P-approx}.
The solid gray curve shows potential $\langle\phi\rangle$ for $q^0_1=q^0_2=0$. (b)-(c) Conductance of
ferroelectric SET vs. parameter $Q_0=-C_g V_g$. The graphs show the hysteresis effect.
Graphs (b) and (c) differ by the direction of $Q_0$ evolution: shown by the arrows. The grey dashed lines correspond to the conductance of SET without ferroelectricity.
Inserts: black lines with arrows show the evolution of potential $\langle\phi\rangle$.
The jump from one branch of $\langle\phi\rangle$ to the other corresponds
to the corresponding vertical lines in the conductance curves.}
\label{fig4}
\end{center}
\end{figure}
For simplicity we consider the case $\delta Q=0$, $V=0$, and $V_s=0$ where
Eq.~\eqref{eqphiap} has a trivial solution $\langle\phi\rangle=0$ and two non-trivial solutions
\begin{gather}\label{1}
\langle\phi\rangle=\pm\frac{E_c}e\left\{\tanh\left(\frac{E_c}T\frac{q^0_1+q^0_2}{e}\right)-
2\frac{q^0_1+q^0_2}{e}\right\}.
\end{gather}
Equation~(\ref{1}) agrees well with numerical results in Fig.~\ref{fig4}
for evolution of average grain potential vs. parameter $Q_0$.
The graph is periodic in $Q_0$ similar to the behavior of average grain potential
of SET in the absence of FE.
However, there are regions in Fig.~\ref{fig4} where parameter $Q_0$ corresponds
to multiple values of average potential $\langle\phi\rangle$.
This behavior appears due to the reorientation of FE polarization by the
average electric field inside the capacitors.
Both FE orientations correspond to the same parameter $Q_0$.
This ambiguity results in hysteresis behavior of the current.

The number of solutions in Eq.~\eqref{eqphiap} depends on the
system parameters $V_s$, $E_c$, and $q_0$. The hysteresis loop shown
in Fig.~\ref{fig4} corresponds to the case of three solutions in Eq.~\eqref{eqphiap}.
The criterion for hysteresis is the following, see Appendix~\ref{Ap3}:
\begin{gather}\label{eqhystcrit}
\frac{q_0}{V_s}\geq \frac{e^2}{E_c}\left(\frac{E_c}{T}-2\right)^{-1}.
\end{gather}
The width of the hysteresis loop is given by the following expression
\begin{gather}\label{eqhystwidth}
\Delta Q_0/2 \approx \frac{q_0}{V_s}\frac{E_c}{e}-\frac{eT}{E_c}.
\end{gather}

\begin{figure}[t]
\begin{center}
\includegraphics[width=1\columnwidth, keepaspectratio]{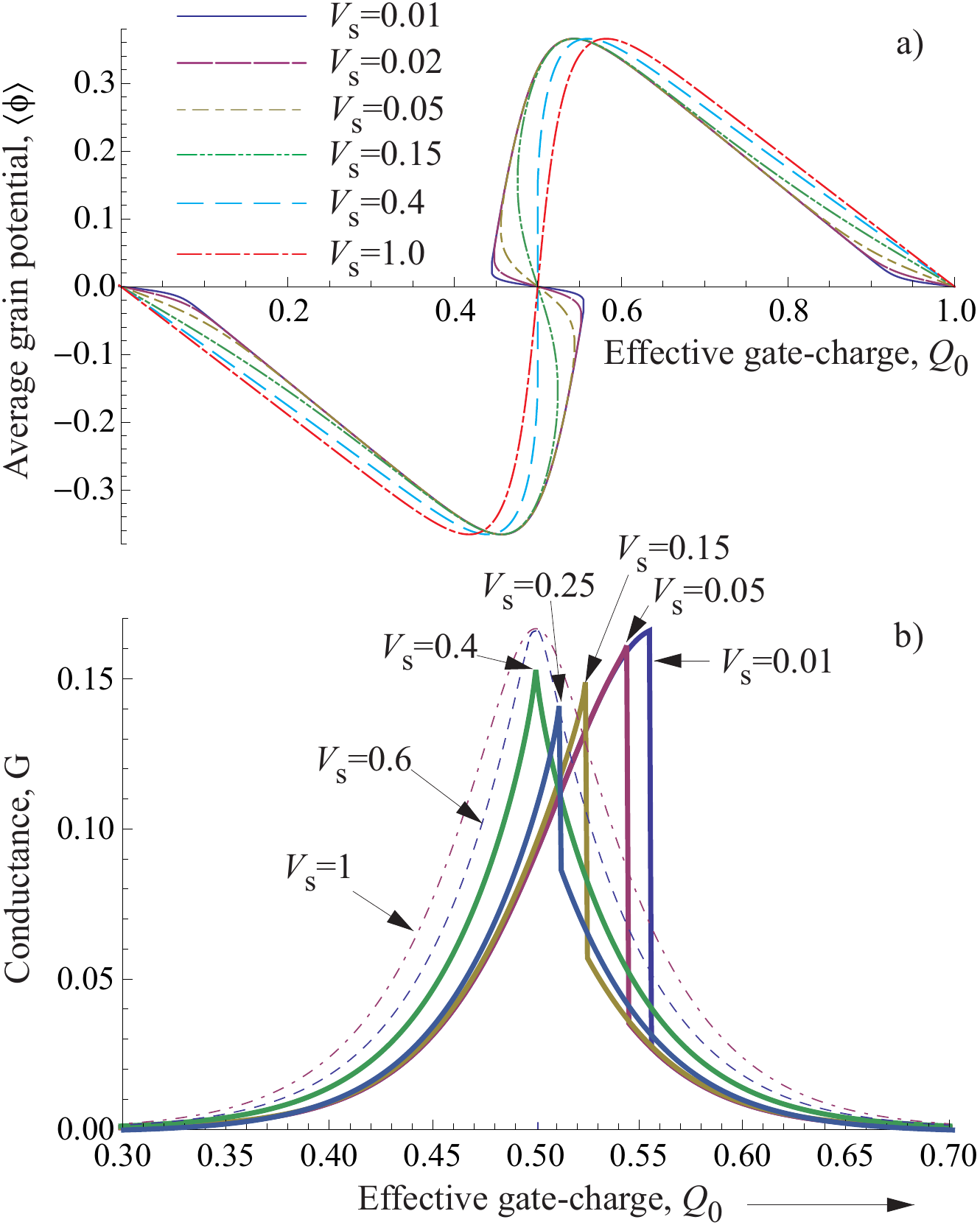}
\caption{(Color online) Average grain potential a) and conductance b) vs. parameter $Q_0$ for different
voltages $V_s$. All parameters (except $V_s$) are same as in Fig.~\ref{fig4}.
The conductance discontinuity responsible for hysteresis becomes smaller with increasing voltage
$V_s$ and completely disappears for voltages exceeding a certain critical value of $V_s$.  }
\label{figVs}
\end{center}
\end{figure}

Fig.~\ref{figVs} shows the change of conductance hysteresis
with voltage $V_s$. It follows that conductance discontinuity
generating the hysteresis decreases with increasing voltage $V_s$ and completely
disappears above a certain critical value of $V_s$, see, e.g., Eq.~\eqref{eqhystcrit}. This result is
natural since increasing voltage $V_s$ produces larger FE
polarizations leading to a more difficult
re-polarization by the external field.

The hysteresis loop is still present even if the
step-like dependence of $q_{\rm fe}$
in Eq.~(\ref{eqphiap}) is substituted by the linear relation $q_{\rm fe}=\alpha_q \langle\phi\rangle$.

Equation~(\ref{eqhystcrit}) can be written using the dielectric
susceptibility of the proper dielectric $\chi_D\approx\alpha_q d/a^2$ as follows $\chi_D>e^2Ta^2/dE_c^2$.
Thus, any dielectric with static susceptibility $\chi_D$ satisfying the above criterion
and with the characteristic reaction time exceeding time $\tau_{e}$ will
produce the hysteresis behavior in the conductivity of SET.
The hysteresis in this model appears due to slow FE (or dielectric).  Then
the FE feels only the average grain potential.
We estimate parameters $q_0$, $V_s$, and the right
hand side of Eq.~\ref{eqhystcrit} in Sec.~\ref{sec:disc}.

The zero voltage conductance of ferroelectric SET vs. $Q_0$ is shown
in Figs.~\ref{fig4}(b)-(c). It is periodic in parameter $Q_0$ similar
to the SET without ferroelectricity. However, the presence of ferroelectricity
breaks the reflection symmetry of conductance peaks and the peaks shape
depends on the direction of $Q_0$ change, see arrows in Fig.~\ref{fig4}(b)-(c).
Therefore there is a hysteresis in the conductance behavior similar
to the branching theory,~\cite{VainbergTrenogin} where the points with
$\frac{d\langle\phi\rangle}{dQ_0}\to\infty$ trigger the jumps between the
different branches of hysteresis loop.

Similar hysteresis behavior shows the conductance density plot
in Figs.~\ref{fig:Diamonds} with Coulomb Diamonds, where Fig.~\ref{fig:Diamonds}(a) and (b)
were obtained with forward and backward change of parameter $Q_0$, while Fig.~\ref{fig:Diamonds}(c) was
obtained for forward-backward evolution of parameter $Q_0$.
Ferroelectricity deforms the Coulomb diamonds: Near the half
integer $Q_0/e$ the Coulomb diamonds acquire the fine structure. However at large enough ferroelectric polarizations this fine-structure can become comparable with the size of the diamonds: The fine-structure characteristic size in the direction of $Q_0$ is $\approx 2[q^0_1+q_2^0]$, for $q^0_1+q_2^0<1/2$.
\begin{figure*}[t]
\begin{center}
\includegraphics[width=1\textwidth, keepaspectratio]{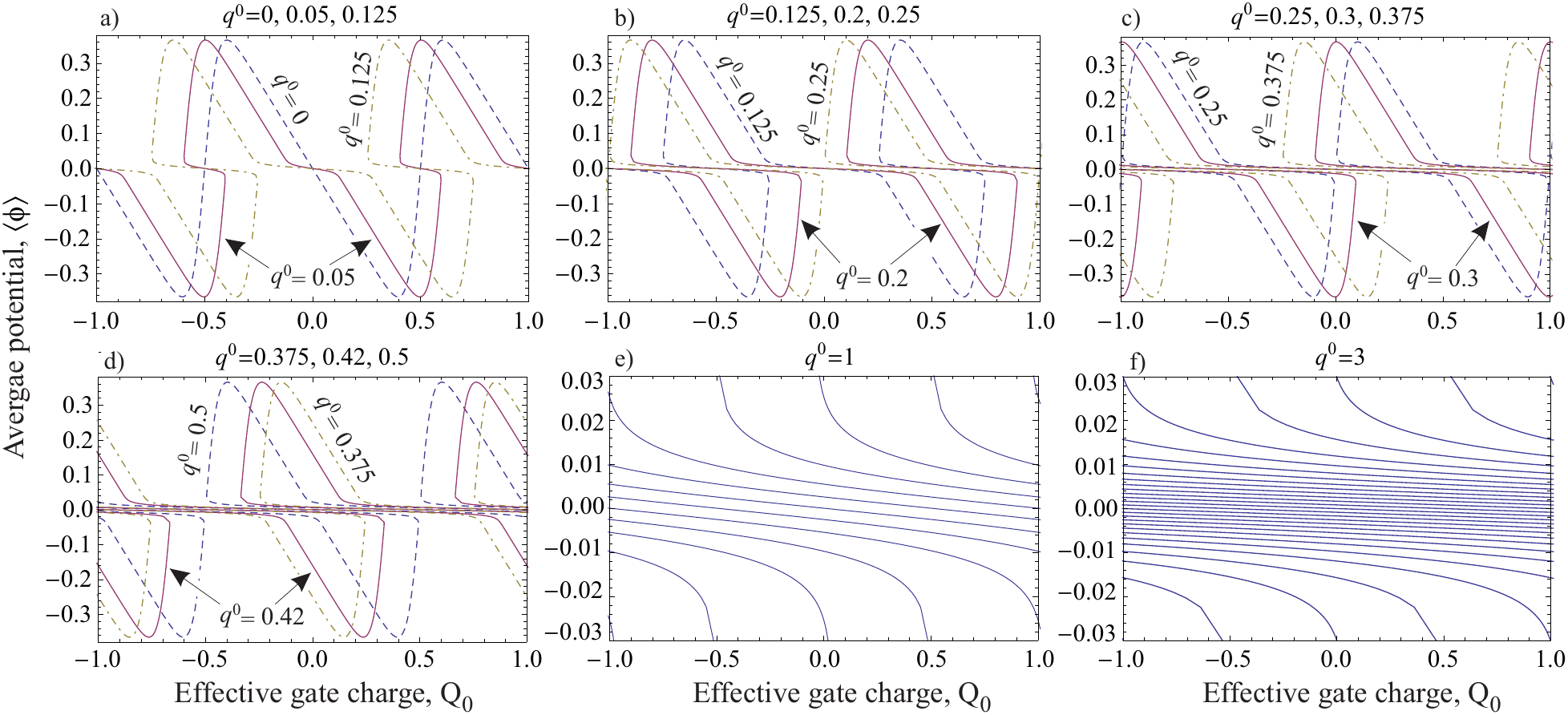}
\caption{(Color online) Dependence of average grain potential  $\langle\phi\rangle$
on parameter $q_1^0=q_2^0=q^0$.
Plots (a)-(f) show the oscillations of
potential $\langle\phi\rangle$ in parameter $q^0$. The shift of potential $\langle\phi\rangle$-peaks induced by the parameter $q^0$
relative to the case $q^0=0$ is $q_1^0+q_2^0=2q^0$. There are two domains of $q^0$ (within the first period) that produce qualitatively different relative positions of  $\langle\phi\rangle$-peaks: (a)-(c) for $0<q^0<1/8$ and $1/4<q^0<3/8$; and (b)-(d) for $1/8<q^0<1/4$ and $3/8<q^0<1/2$. As follows from (e)-(f) the small-scale branch of $\langle\phi\rangle$ is non-periodic in parameter
$q_0$, but it is periodic in  $Q_0$. }
\label{figQ0}
\end{center}
\end{figure*}

\begin{figure}[tb]
\begin{center}
\includegraphics[width=0.89\columnwidth, keepaspectratio]{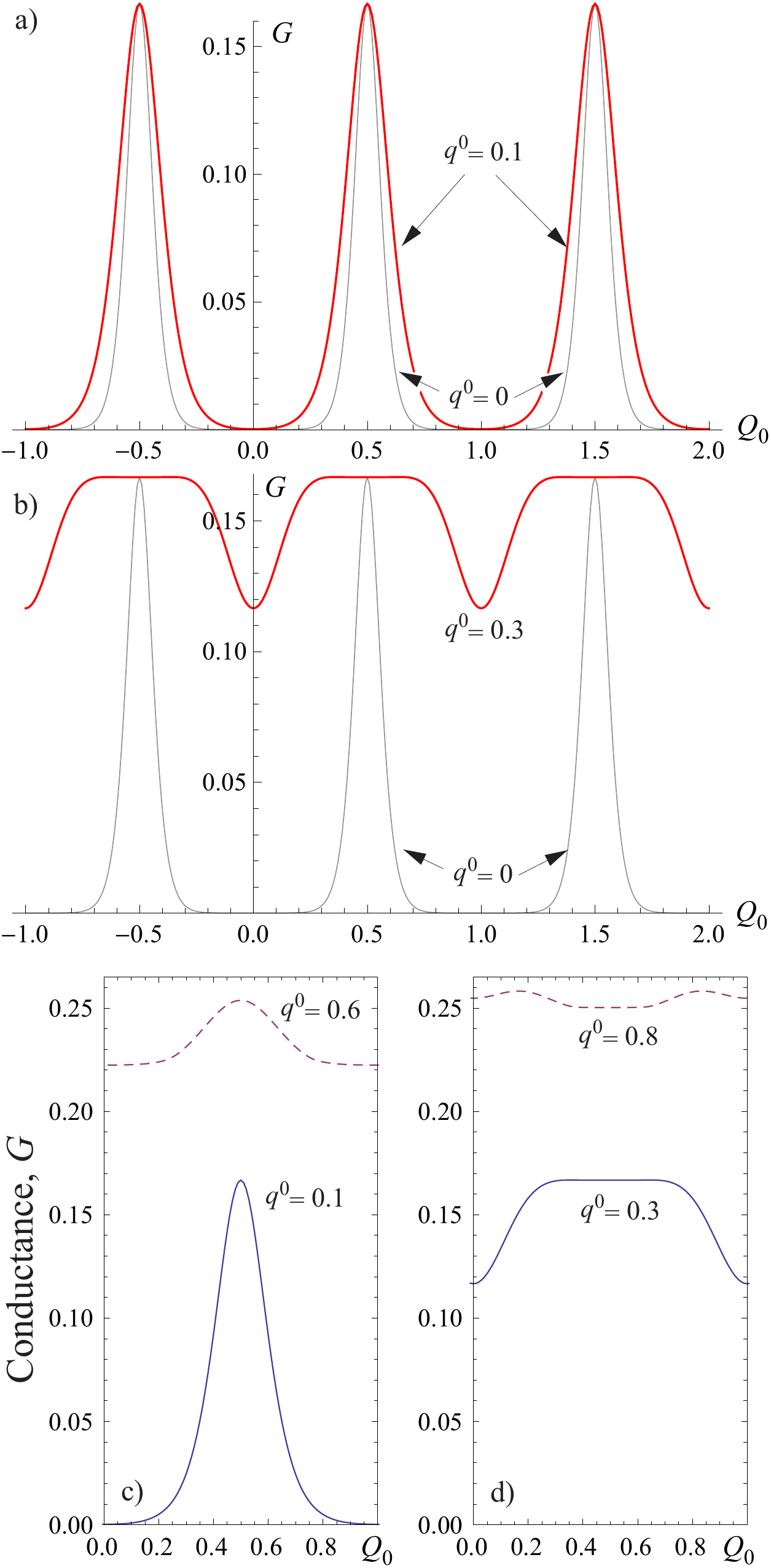}
\caption{(Color online) Graphs (a) and (b) show the broadening of conductance peaks
due to ferroelectricity: Red graph in (a) corresponds to $q^0_1=q^0_2=0.1$, while in (b)  $q^0_1=q^0_2=0.3$.
The grey graphs show the peaks for $q^0_1=q^0_2=0$. Graphs (c) and (d) show the
reduction of the peaks amplitude with increasing $q^0$.
The step of the $q^0$ increase is $0.5$ similar to the``period'' in Fig.~\ref{figQ0}.
For plots (a)-(d) we use the following set of parameters:
$T = 0.03$, $C_1 = 0.3$, $C_2 = 0.5$, $C_g = 0.2$, and $R_2=2R_1$, like in Fig.~\ref{fig4}.}
\label{fig:f1}
\end{center}
\end{figure}

The hysteresis can be better understood using the energy balance consideration.
The effective free energy of SET with $n$ excess charges on the grain
for zero temperature and bias voltage $V$ has the form
\begin{gather}\label{eqF}
  F=E_c\min_n\left(n-\left(Q_0+q_{\rm fe}\right)/e\right)^2.
\end{gather}
Below we use dimensionless units discussing Eq.~\eqref{eqF}. First, we compare
the energies of the system for $Q_0=1/2$.
In this case for average grain potential $\langle\phi\rangle$
according to Fig.~\ref{fig4} three choices are possible: $\langle\phi\rangle=0$ and $\langle\phi\rangle=\pm \phi_0$, where $\phi_0\approx 0.4\gg V_s$.  The first choice corresponds to $q_{\rm fe}=0$ while two other choices to $q_{\rm fe}\approx \pm 2q^0$.
The solution with $\langle\phi\rangle=0$ corresponds to $F/E_c=1/4$. [This value corresponds to
the crossing point, $Q_0=1/2$, of two parabolas, $\left(n-Q_0\right)^2$, $n=0,1$ as
functions of $Q_0$.] For two other cases the free energy, $F/E_c$, is smaller by $2q^0 [1-2q_0]$.
Here we choose  $q^0 < 1/2$, thus the minimum in Eq.~\eqref{eqF} corresponds to $n=0$ or $n=1$.
The solution $\langle\phi\rangle=0$ is physically unstable at $Q_0=1/2$ since it has
the largest free energy. Similar consideration can be used in explaining the
jumps between different branches of $\langle\phi\rangle$ in Figs.~\ref{fig4}(b)-(c).

Figure~\ref{fig4} shows that at zero voltage one can
drive the system between two states with FE layers polarized toward or backward directions
with respect the grain by changing parameter $Q_0$. This behaviour can be understood as follows:
At zero bias voltage there is no preferable direction in the SET.
Contrary, a finite bias voltage results in
electric field which breaks the symmetry of the problem
leading to two FE polarizations in parallel.

We confirm this presenting numerical calculations of FE polarizations in
the $(Q_0,V)$-plane, Fig.~\ref{figPhaseDiag}, where the color gradients and the arrows
indicate the polarizations of the left and the right ferroelectrics.
Plots (a) and (b) show FE polarizations for increasing parameter $Q_0$
[similar to Figs.~\ref{fig:Diamonds}a and \ref{fig4}b] while plots (e) and (f) show this polarization
for decreasing $Q_0$ [similar to Figs.~\ref{fig:Diamonds}b and \ref{fig4}c].
In fact, these graphs show the charges in the grain that screen
the FE polarization. Graphs (c) and (d) show the arithmetical mean of the polarizations
corresponding to the left and to the right ferroelectrics.
To distinguish the non-zero total screening charge in the parallel case we choose parameters in
Fig.~\ref{figPhaseDiag} slightly different from Figs.~\ref{fig:Diamonds}-\ref{fig4}: $q_1^0=0.03$ and $q_2^0=0.06$.

Figure~\ref{figQ0} shows the evolution of average grain potential $\langle \phi\rangle$ for $q_1^0=q_2^0=q^0$.
There are several branches in the behavior of $\langle\phi\rangle$ depending on the ratio $\langle\phi\rangle/V_s$.
The peaks correspond to the first branch. The nearly linear segments of $\langle\phi\rangle$ with the maximum much smaller than the peak hight correspond to the second branch. Figure~\ref{figQ0}(a)-(c) shows
that the peaks of $\langle\phi\rangle$ are periodic over $q^0$ with the period of  $0.5$~($|e|$). The shift of $\langle\phi\rangle$-peaks
at $q^0>0$ relative to the case $q^0=0$ is $q_1^0+q_2^0=2q^0$. The terms
with $q^0_i$, $i=1,2$ enter the expression for potential $\langle\phi\rangle$ similar
to the shift-renormalization of parameter $Q_0$. Figure~\ref{figQ0}(a)-(d) and
(e)-(f) show that the second branch of potential $\langle\phi\rangle$ is strongly non-periodic.

\subsection{Fast ferroelectric. Polarization follows charging-discharging events.\label{Sec2}}

Now we consider the opposite case of fast polarization following the charging-discharging process,
$\tau_{\rm\scriptscriptstyle P} \lesssim \tau_e$. In this limit the polarization $P$ depends
on the instant electric field $\mathcal E(n)$ instead of the average electric field $\langle\mathcal E\rangle$ as it was
discussed before.
Here we replace $Q_0$ in the orthodox theory by $Q_s= Q_0+q^0_1\tanh\left(\frac{\phi(n)+V/2}{V_s}\right)+ q^0_2\tanh\left(\frac{\phi(n)-V/2}{V_s}\right)$, Appendix~\ref{Ap1}. With this replacement
Eqs.~\eqref{eqphia}-\eqref{qfe} remain valid with substitution of potential
$\phi(n)$ instead of average potential $\langle \phi\rangle$ in Eq.~\eqref{eqphia}.

The conductance behavior is shown in Fig.~\ref{fig:f1}. Ferroelectricity
preserves periodicity over the parameter $Q_0$ similar to the mean-field theory discussed
in Sec.~\ref{Sec1}. However, in this limit the hysteresis is absent while
the broadening of the conductance peaks, Figs.~\ref{fig:f1}(a)-(b), and the reduction of the
peaks amplitude with increasing $q^0$ are present, Figs.~\ref{fig:f1}(c)-(d).

In orthodox theory the conductance of SET in the absence of ferroelectricity and at low temperatures, $T\ll E_c$
follows the following relation
\begin{gather}\label{go}
G(\delta Q_0)=\frac 12 \cdot \frac 1{R_1+R_2} \cdot \frac{e\,\delta Q_0/\Cs T}{\sinh(e\,\delta Q_0/\Cs T)},
\end{gather}
where $\delta Q_0 = \min_k[Q_0 - (2k+1)\frac e2] \ll e$ is the deviation from the
degeneracy point. The width of conductance peaks defines the temperature-parameter $T/E_c$.

For FE the degeneracy points do not follow exactly the half integer $Q_0/e$.
Above it was shown that FE polarization redefines $Q_0 \to Q_s$ where parameter $Q_s$
depends on the polarization and the excess charge number $n$.
Therefore the conductance peak in Fig.~\ref{fig:f1}(a)-(b) has the width
$(q^0_1+q^0_2)/2$ and consists of many shifted conductance peaks \eqref{go}.
Thus the width of the peak-plato in Fig.~\ref{fig:f1}b is approximately $(q^0_1+q^0_2)/2=|e|/3$.
Similar arguments explain the reduction of the conductance peaks
amplitude with increasing parameter $q^0$ in Figs.~\ref{fig:f1}(c)-(d).

The question about the average direction of polarizations can be investigated similar
to the previous section. The results are similar, but in this case the hysteresis is absent.

At large $V_s$ performing the linear expansion in electric field/voltage in Eqs.~\eqref{P-approx}-\eqref{eqphia} we reproduce the result of orthodox theory for
potential $\phi(n)$ with renormalized capacitances, $C_i\to C_i +q_i^0/V_s^{(i)}$.
Therefore for zero-field differential dielectric susceptibility of the capacitor-$i$ we find
$\epsilon^{(i)}=1+q_i^0/C_i V_s^{(i)}$.

\section{Discussion \label{sec:disc}}

\subsection{Ferroelectric model}

\begin{figure}[t]
  \centering
  \includegraphics[width=0.99\columnwidth]{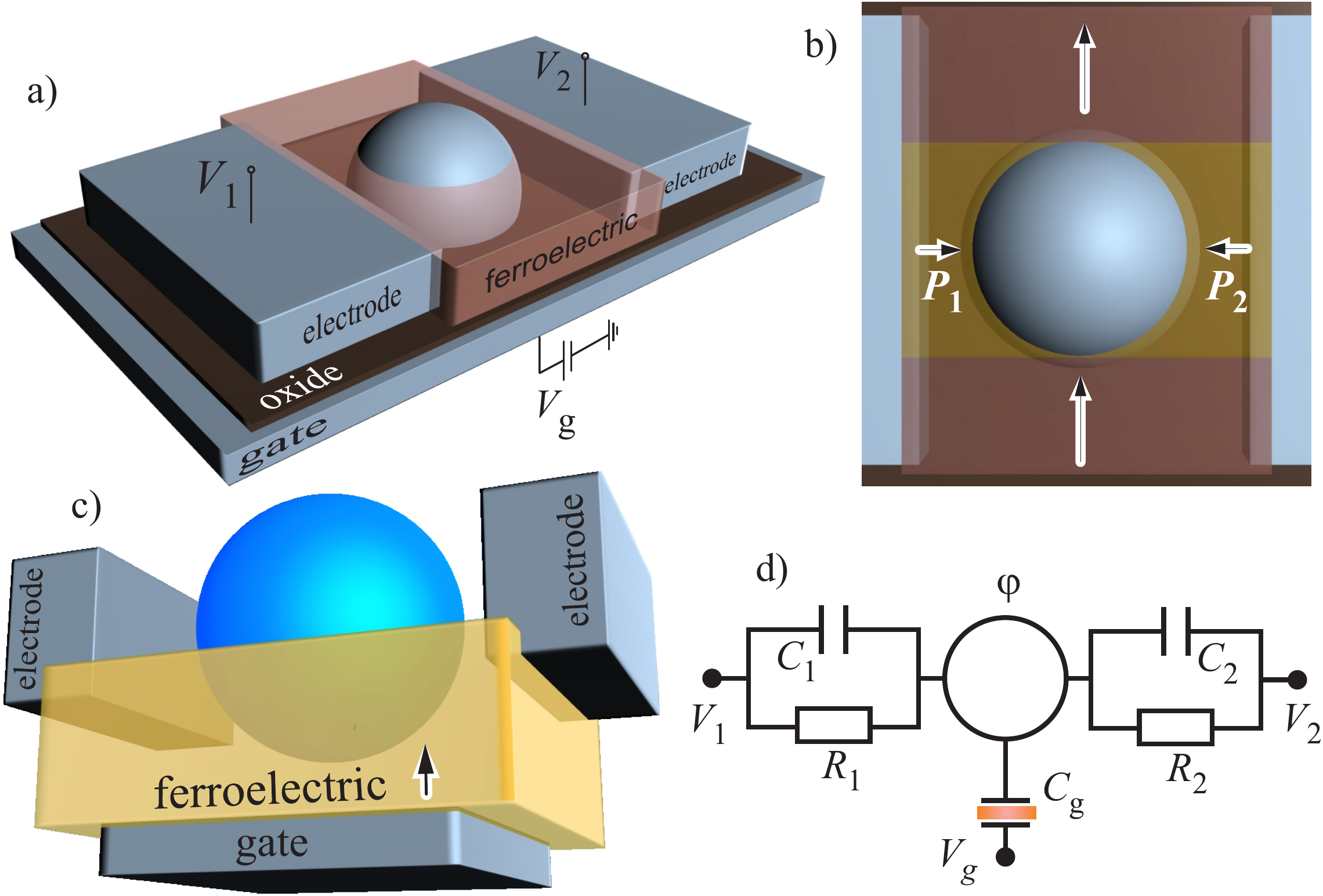}\\
  \caption{(Color online) a) Possible experimental setup with nanograin placed in a bulk ferroelectric material.
  b) Top view. In the bottleneck between the electrodes and the grain, highlighted by the orange color,
  the ferroelectric layer is thin (quasi-two-dimensional) with polarization being
  unbind from the bulk ferroelectric. c) Different geometry: Ferroelectric placed between the grain and the gate.
  In this case there is no restriction on the thickness of the layer: For SET
  no tunneling is required between the gate and the grain. d) The equivalent electrical scheme of the SET-device shown in (c). }\label{fig1old}
\end{figure}

Ferroelecric SET consists of nanosized charged metallic grain embedded in a ferroelectric
confined by the metallic leads, Fig.~\ref{fig1old}(a).
The thickness of FE layer between the grain and the leads is few nm.
It is known that even for such a thin FE film the continuum theories
of ferroelectricity are valid.~\cite{ahn2004ferroelectricity,Lukyanchuk2005modeling} To determine the state of FE under the influence of the charged grain one needs to solve the inhomogeneous
Landau-Ginzburg-Devonshire (LGD) equation.~\cite{Levan1983} This question appears
frequently in problems dealing with local modification of FE properties by the
tip of scanning probe microscope.~\cite{Rodriguez2010} We assume that domain wall
thickness $l_{\rm d}$ in FE is less than the grain size, $l_{\rm d}\ll a$.~\cite{Rodriguez2010,Kalinin2009}
The grain influences only the FE region between its surface and the leads, Fig.~\ref{fig1old}(b).
Inside this region polarization is homogeneous and depends on the grain state.
Outside this region the FE state does not depend on the grain charge. The side regions do not affect the
electron transport.~\cite{Udalov2014PRB89} With these assumptions the
homogeneous LGD theory is valid for the description of FE behavior.

The FE material can be placed not only between the grain and the leads but also between the grain and
the gate electrode, Figs.~\ref{fig1old}(c)-(d). In this case the
FE layer does not have any metallic inclusion and can be made as a rather
thick film. Such a geometry is relevant for experiment and
allows to avoid problems with the influence of grain shape
on the FE polarization. The transport equations for such SET are
similar to the transport equation written above. For example, in Eq.~\eqref{qfe}
one should use $q_{\rm fe} =q^0\tanh\left(\frac{\langle\phi\rangle-V_g}{V_s}\right)$,
with $q^0$ being related to the FE between the gate and the grain.

The memory effect (hysteresis) in ferroelectric SET can be used for computer memory-cell
with the measurement of the zero-bias conductance being the reading operation while the
application of the gate voltage being the writing operation. Such a memory-cell will be discussed in the
forthcoming publication.

\subsection{Evaluation of parameters}

In this section we discuss important physical parameters of FE SET such as FE and SET time
scales, electric field due to metallic grain, the FE switching field,
and the FE saturation polarization. These parameters define the physical behavior of FE SET.

In the previous sections we discuss two limits: i) slow ($\tau_e<\tau_{\rm\scriptscriptstyle P}$)
and ii) fast ($\tau_e>\tau_{\rm\scriptscriptstyle P}$) ferroelectric. Estimates show that the
characteristic time $\tau_e=\Rs\Cs$ varies in a rather large range from dozens of nano- to picoseconds.
This time is controlled by the system geometry and materials. The distance between the
grain and the leads controls the resistivity of the SET, $\Rs$, the dielectric
properties of the FE material, and the capacitance of the SET, $\Cs$.
The FE switching time $\tau_{\rm\scriptscriptstyle P}$ depends on the material and
can be in the range of $10^{-6}$~s~\cite{Krupanidhi1993} to few nanoseconds.~\cite{Cuppens1991}
Therefore both limits are relevant for experiment. SET with changing energy $E_c \sim 300$~K have
small capacitance, $\lesssim 10^{-17}$~F leading to $\tau_e\ll\tau_{\rm\scriptscriptstyle P}$.

Discussing  two limits we neglect the hysteresis loop of FE material (and thus, the spontaneous polarization). This assumption is valid
for large electric field created by a single electron in a grain in comparison with
the FE switching field, $\mathcal E_{\rm el}\gg \mathcal E_s$. This is typical for
number of FE including Li-doped ZnO,~\cite{Liu2003} Pb(In$_{1/2}$Nb$_{1/2}$)$_{1-x}$Ti$_x$O$_3$,~\cite{Tseng2007}  (PbMg$_{1/3}$Nb$_{2/3}$O$_3$)$_x$(PbTi)$_3$)$_{1-x}$,~\cite{Schmidt2002} PZT,~\cite{Krupanidhi1993PZT} and  etc.
The presence of hysteresis loop leads to more complicate picture of electron transport in
FE SET with the interplay of FE hysteresis loop and the hysteresis appearing due to the
interaction of FE with the grain, Sec.~\ref{Sec1}. In the opposite limit, $\mathcal E_{\rm el}\ll \mathcal E_s$,
the polarization becomes an external parameter
as in the ordinary FE tunnel junctions.

The magnitude of FE saturation polarization strongly affects the
electron transport for fast ferroelectrics, $\tau_e>\tau_{\rm\scriptscriptstyle P}$.
If induced charge due to FE exceeds one electron charge the Coulomb blockade is suppressed
leading to the conductivity independent of gate voltage.
In this case the FE completely screens the electric field of an electron
on the grain. To observe the conductivity peaks in Fig.~\ref{fig:f1} the FE environment
should generate the charge smaller than one electron, ($q^0_1+q^0_2<|e|$). Typical ferroelectrics,
such as P(VDF-TrFE), PZT, (PbMg$_{1/3}$Nb$_{2/3}$O$_3$)$_x$(PbTi)$_3$)$_{1-x}$, have bulk
polarization about $P=1$~e/nm$^2$ leading to $q^0_i\gg |e|$, $i=1,2$ for a few nm size grain.
However, decreasing the thickness of FE film reduces its polarization.~\cite{Frid2006rev} For example, drastic polarization reduction from 1.5 e/nm to 0 e/nm is predicted for BaTiO$_3$ when the thickness of the BaTiO$_3$ film decreases from 15 nm to 3 nm.~\cite{junquera2003Nature} Suppressing of polarization with decreasing of FE thickness was observed in P(VDF-TrFE) films.~\cite{zhang2001JAP}

For slow FE, Eq.~\ref{eqhystcrit} separates two regimes of FE SET
with finite and zero hysteresis conductivity voltage dependencies.
For estimates we write voltage $V_s$ using FE switching field, $V_s=d\cdot \mathcal E_s$
and the charge $q^0$ [we drop index $i$] using the FE polarization, $q^0\approx P\cdot a^2$. For $P\approx 0.05$~e/nm,~\cite{Liu2003} $d\approx2$~nm, $a\approx 3$~nm, and temperature $T\approx 100$~K we find the criterion for the
appearance of hysteresis, $\mathcal E_s<10$~MV/cm.
This criterion is valid for almost all ferroelectrics. In addition, we note that in the case of slow FE the condition ($q^0_1+q^0_2<|e|/2$) results in simple hysteresis behavior of the system, its violation makes the behavior more complicated, but does not affect the existence of hysteresis.

\section{Conclusion}\label{sec:Conclusion}
We investigated the electron transport in single-electron-device
with ferroelectric active layers. We showed that there is an interplay of
ferroelectricity and single-electron tunneling. We distinguish two different cases of slow and fast ferroelectric. In the first case the
gate voltage dependent conductance shows the instability related to the spontaneous polarization inversion of ferroelectric polarizations. We show that similar instability may show also SET with slow dielectric. At small bias voltages the polarization can be switched by a single excess electron in the grain.
In the case of fast ferroelectric instability is absent. However, we show that the Coulomb-Blockade peaks of zero-bias conductance as the function of the gate-voltage become wider and finally disappear with increasing of the FE polarizations. Such an effect appears due to strong non-linear screening of electron charge in the grain by ferroelectrics leading to the suppression of the Coulomb Blockade. Finally we show that our results could be observed experimentally.

\section{Acknowledgments}

N.~C. was partly supported by RFBR No.~13-02-00579, the Grant of President of Russian
Federation for support of Leading Scientific Schools, RAS presidium and Russian Federal Government programs.
I.~B. was supported by NSF under Cooperative Agreement Award EEC-1160504, NSF Award DMR-1158666,
and NSF PREM Award..

\appendix
\section{Orthodox theory of Ferroelectric Single Electron Transistor \label{Ap1}}

\subsection{Main equations}

Below we outline the main steps that help to understand our results in the presence
of ferroelectricity using the language of orthodox theory

In orthodox theory the rate describing the change of grain charge from $n$ to $n+1$ electrons
through the first tunnel barrier, the left one in Fig.~\ref{fig2}a), is
\begin{equation} \label{eq:Gamma_1f}
    \Gamma^{(1)}_{n\to n\pm1}=\frac1{e^2 R_1}\cdot \Delta F_1^\pm N_B\left( \Delta F_1^\pm \right),
\end{equation}
where $N_B(\omega)=1/[\exp(\omega/T)-1]$ is the Bose-function and $R_1$ is the tunnelling
bare resistance. Similar expression can be written for the discharge process
through the second tunnel barrier by changing the index ``1'' to ``2''.
Here $ \Delta F_1^\pm$ is the change of effective free energy between the initial and the
final states
\begin{subequations} \label{eq:dF}
\begin{align} \label{eq:dF1}
    \Delta F_1^\pm &=\frac{e^2}{\Cs}\left\{ \frac12\pm\left(n-\frac{Q_s}e \right) \pm \frac{(C_2+C_g/2) V}e\right\},
    \\
    \Delta F_2^\pm &=\frac{e^2}{\Cs}\left\{ \frac12\pm\left(n-\frac{Q_s}e \right) \mp \frac{(C_1+C_g/2) V}e\right\}.
\end{align}
\end{subequations}
where
\begin{gather}
 Q_s= Q_0+\sum_i\int_i d\mathbf n_i\cdot \mathbf P_i.
\end{gather}
In the orthodox theory, $Q_s= Q_0$. For ``slow'' ferroelectric we have $\int_i d\mathbf n_i\cdot \mathbf P_i=q_{\rm fe} =q^0_1\tanh\left(\frac{\langle\phi\rangle+\frac{V}{2}}{V_s}\right)+ q^0_2\tanh\left(\frac{\langle\phi\rangle-\frac{V}{2}}{V_s}\right)$,
while for ``fast'' we find
$\int_i d\mathbf n_i\cdot \mathbf P_i=q^0_1\tanh\left(\frac{\phi(n)+\frac{V}{2}}{V_s}\right)+ q^0_2\tanh\left(\frac{\phi(n)-\frac{V}{2}}{V_s}\right)$.
The $\Gamma_{n\to n+1}$--rates in the detailed-balance relations~\ref{eq:ballance} are defined as follows
\begin{subequations} \label{Gto}
\begin{align}
\Gamma_{n\to n+1} &\equiv \Gamma^{(1)}_{n\to n+1} +\Gamma^{(2)}_{n\to n+1}, \\
\Gamma_{n\to n-1} &\equiv \Gamma^{(1)}_{n\to n-1} +\Gamma^{(2)}_{n\to n-1}.
\end{align}
\end{subequations}

\subsection{Approximation near the ``degeneracy point''\label{Ap2}}

The probabilities near the degeneracy point $Q_0=1/2$ can be found using the orthodox theory
\begin{gather}\label{po1}
  p(0)=\frac{\Gamma_{1\to 0}}{\Gamma_{0\to 1}+\Gamma_{1\to 0}}, \qquad p(1)=\frac{\Gamma_{0\to 1}}{\Gamma_{0\to 1}+\Gamma_{1\to 0}}.
\end{gather}

Here Eq.~\eqref{eq:dF} reduces to
\begin{gather}
\Delta F_{1}^{+}(0)=-2\frac{E_c}{e}\left[\delta Q_s-\left(C_2+\frac{C_g}2\right)V\right],
\\
\Delta F_{2}^{+}(0)=-2\frac{E_c}{e}\left[\delta Q_s+\left(C_1+\frac{C_g}2\right)V\right],
\end{gather}
where $\delta Q_s= \delta Q_0+\sum_i\int_i d\mathbf n_i\cdot \mathbf P_i$ and $\delta Q_0=Q_0-e/2$. Here $\Delta F_{1}^{+}(0)=-\Delta F_{1}^{-}(1)$ and $\Delta F_{2}^{+}(0)=-\Delta F_{2}^{-}(1)$. Using Eqs.~\eqref{Gto} and \eqref{po1} we find, $p(1)^{-1}={1+\exp[2 E_c (\delta Q_0+q_{\rm fe})/ eT]}$. Finally, using Eq.~\eqref{eq_phi_av} we obtain Eq.~\eqref{eqphiap}.

\subsection{Hysteresis width\label{Ap3}}

The criterion for conductivity hysteresis in Eq.~\ref{eqhystcrit} can be derived using Eq.~\ref{eqphiap}
for average potential. This equation has three solutions if the slope, derivative of $\langle\phi\rangle$,
of the function in the right hand side is larger than the slope of the linear dependence in the left hand side.

The estimate of the hysteresis width, $\Delta Q$, can be done using the following assumptions:
1) the polarization is linearly depend on the average
potential $q_{\mathrm{fe}}(\langle\phi\rangle)\approx q_0/V_s$; 2) we replace the hyperbolic tangents
by the piecewise straight function, $Z(x)=1$ if $|x|>1$ and $Z(x)=x$ if $|x|<1$; and 3) we neglect the
``slow'' function $2q_{\mathrm{fe}}(\langle\phi\rangle)/e$ in the right hand side.

For $E_c/T\gg1$ the criterion of the conductivity hysteresis, Eq.~\ref{eqhystcrit}, can be
obtained using the formula for hysteresis width considering $\Delta Q>0$.

\bibliography{our_bib}

\end{document}